# How and When do Planets Form?
# The Inner Regions of Planet Forming Disks at High Spatial and Spectral Resolution

**Rafael Millan-Gabet**

California Institute of Technology
NASA Exoplanet Science Institute
(626) 395-1928
R.Millan-Gabet@caltech.edu

**John D. Monnier**

University of Michigan
Astronomy Department
(734) 763-5822
monnier@umich.edu

*With contributions from:*

**Rachel Akeson** (Caltech/NExSci)
**Chas Beichman** (Caltech/NExSci)
**Sean Brittain** (Clemson Univ.)
**Theo ten Brummelaar** (Georgia State Univ.)
**Nuria Calvet** (Univ. Michigan)
**Josh Eisner** (Univ. Arizona)
**Phil Hinz** (Univ. Arizona)
**Hanna Jang-Condell** (Univ. Maryland)
**Marc Kuchner** (Goddard Space Flight Center)
**Fabien Malbet** (LAOG, France)
**Sean Matt** (Ames Research Center)
**Joan Najita** (National Optical Astronomy Observatory)
**Sean Raymond** (Univ. Colorado)
**Aki Roberge** (Goddard Space Flight Center)
**Ajay Tannirkulam** (Univ. Michigan)
**Neal Turner** (Jet Propulsion Laboratory)
**David Wilner** (Harvard-Smithsonian Center for Astrophysics)

## Summary

The formation of planets is one of the major unsolved problems in modern astrophysics. Planets are believed to form out of the material in circumstellar disks known to exist around young stars, and which are a by-product of the star formation process. Therefore, the physical conditions in these disks - structure and composition as a function of stellocentric radius and vertical height, density and temperature profiles of each component - represent the initial conditions under which planets form. Clearly, a good understanding of disk structure and its time evolution are crucial to understanding planet formation, the evolution of young planetary systems (e.g. migration), and the recently discovered, and unanticipated, diversity of planetary architectures. However, the *inner* disk regions (interior to ~10 AU) most relevant in the context of planet formation are very poorly known, primarily because of observational challenges in spatially resolving this region.

Recent advances in milliarcsecond imaging capabilities, and further developments underway for the next decade, will enable direct and fundamentally new measurements of the inner disk: the radial location of the sources of continuum and line emission, gas chemistry and dust mineralogy, and surface brightness. Combined with high resolution spectroscopy and realistic modeling of the relevant phenomena, we can derive physical properties of the inner disk that are crucial to understanding planet formation and evolution: temperature and density profiles, accretion rates and viscosity ($\alpha$) parameter. We identify the following fundamental questions that these methods can help to answer in the next decade:

1. What is the dominant mode of planet formation? Measurements of young disks can determine whether physical conditions best support gravitational instability or core-accretion models. Can the differences between exoplanet architecture observed as a function of host star mass be explained by differences in the disk initial conditions? Do "dead zones" exist in circumstellar disks and what are their consequences?

2. What mechanisms are responsible for transporting angular momentum in the inner disk? Removal of angular momentum is key to understanding the origins not only of the stars but also of the planets, since it determines the rate at which material spirals in to accrete on the young star. New data on disk-locking and rotational evolution are not well-supported by the current paradigm, and the nature of jets and disk winds have remained uncertain for years. New high-resolution spectral and spatial information can elucidate the star-disk connection.

3. When do planets form? In the previous decade, we discovered that disks "disappear" within a few million years and that young giant planets interact with gaseous circumstellar disks, driving migration. Over the next decade, interferometric imaging can spot gaps, over-densities, or even the accreting proto-planets themselves.

## Introduction and Context

Figure 1 shows an artistic rendition of a young pre-planetary disk around a solar mass star. We have indicated on it the most relevant physical processes and their associated spatial scales. While the last two decades have seen fundamental advances in many aspects of our understanding of disks around young stars (e.g., recent reviews in Protostars and Planets V), most have addressed the "outer" disk, at 10s – 1000s AU from the central star.

However, it is the inner disk regions, within ~10 AU from the central star, which are most relevant to the planet formation problem. In our Solar System, all the rocky planets are located within 2 AU of the Sun. While giant gas planets are believed to form further out ($\geq 5$ AU), extra-solar hot-Jupiters are commonly found at a fraction of an AU from their host stars - likely as a result of migration through the inner disk. The dust evaporation front, habitable zone, and snowline are all located within $0.1 - 10$ AU, depending on stellar mass.

The terrestrial planet forming region itself can be probed using high resolution spectroscopy of infrared lines (Najita et al. 2007a) both molecular (e.g. CO, OH, H2O) and atomic (e.g. OI, NeII). The kinematics of these lines pinpoint their location within a few AU of the central star, and a disk velocity field (e.g. Keplerian) can be assumed to infer the radii from which the emission arises. However the interpretation is made difficult by the necessity to make simplistic assumptions regarding the geometry as well as temperature and density distributions of these inner regions. While much progress has been made, we have only a crude understanding of the true conditions in the inner disk, due to a lack of high angular resolution observations.

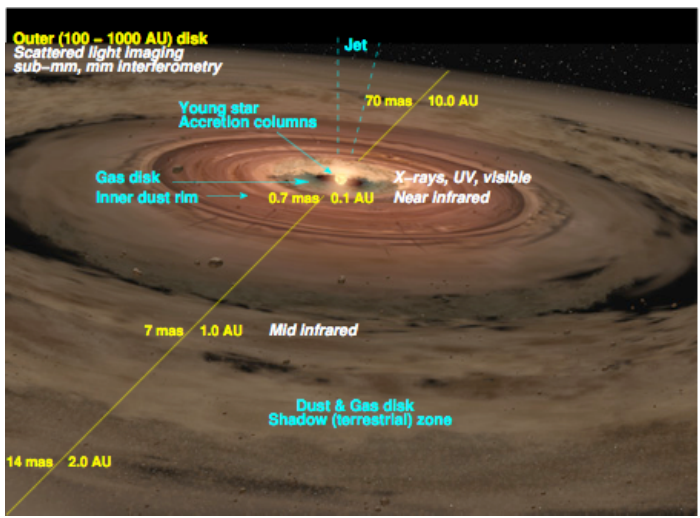

**Figure 1:** Representation of a young pre-planetary disk around a solar mass star. The angular scales correspond to a distance of 140 pc (e.g. Taurus). This contribution focuses on establishing the physical conditions in the inner disk (< 10 AU) in order to determine the initial conditions for planet formation. The inner disk cannot be spatially resolved by traditional imaging techniques, but is resolvable using long baseline infrared interferometry (Credit: based on artist's conception by David Darling).

For example, even at the distance of one of the nearest star forming regions (e.g. 140 pc for Taurus) spatial scales $\leq 1$ AU correspond to angular scales $\leq 7$ milliarcsecond (mas). The characteristic temperatures in these regions imply that emission is at visible to infrared wavelengths, and no existing or planned telescope (even the next generation of $20 - 30$m telescopes) has the required diffraction-limited resolution. However, long-baseline interferometers operating in the infrared can synthesize the resolving power of a large aperture using two or more widely-separated telescopes – an angular resolution of $\sim 1$ mas is routinely achieved today.

Using these techniques, significant first steps have been taken in the last ten years that have already altered our view of the inner regions of planet forming disks. As was reviewed in the most recent Protostars & Planets volume (Millan-Gabet et al. 2007, and references therein) we may list the following as the principal contributions: **(1)** The near infrared characteristic disk sizes have been measured, and from these measurements it has been inferred that the dust evaporation front is located at $0.1 - 1$ AU for young stars in the mass range $1 - 4$ $M_{sun}$, and that the gas interior to it is largely optically thin. These findings essentially rule out popular geometrically-thin, optically-thick disk models (Shakura & Sunyaev 1973), and have motivated a new class of models that pay particular attention to the detailed location and shape of the inner dust rim and shadow region; **(2)** the basic disk morphology has been established, ending for example the "disk vs. envelope" debate of the 1990s; **(3)** the characteristic sizes and shapes of the mid-infrared disk have also been measured, as well as radial gradients in dust mineralogy.

These pioneering efforts may however be characterized as being based on: **(a)** sparse data (typically single baseline data with limited spatial frequency coverage and limited to a single broadband spectral channel); **(b)** interpretation dependent on relatively simple modeling (typically assuming reasonable geometric approximations to the disk brightness, and deriving fundamental parameters such as characteristic sizes and shapes); and **(c)** small sample sizes (limited by sensitivity). They also tended to concentrate on the dust component, understandably since it dominates the opacity and infrared emission.

## Spatially and Spectrally Resolving the Inner Disk

Although the dust dominates the infrared emission and opacity, gas dominates the inner disk mass and dynamics; and is expected to exist even inside the dust sublimation radius (ultimately accreting onto the young star). The ambiguities in interpreting the high resolution spectroscopy data mentioned above can be lifted by interferometer data that spatially resolves the emission and is also spectrally resolved. First steps have also recently been taken using this technique, and the principle is illustrated in Figure 2. Indeed, hot gas has been found in some cases to be spatially located interior to the dust sublimation radius.

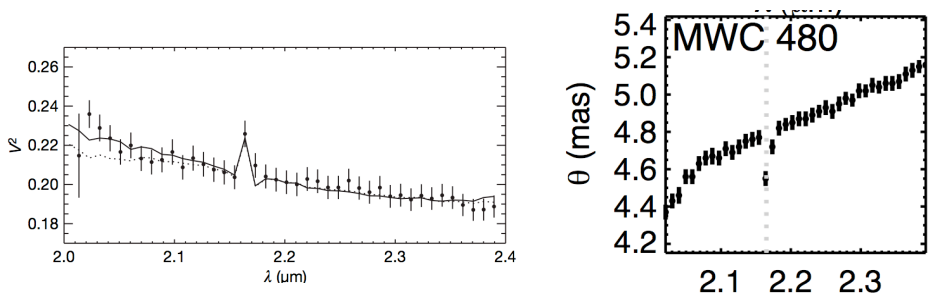

**Figure 2:** Spectro-interferometric measurements of the young stellar object MWC 480 (R=230) from Eisner 2007 & 2009. Wavelength dependence of the fringe visibility (left) and equivalent size of the near-infrared emitting region (right). Measurements such as these provide a powerful probe of the disk profile (see also Kraus et al. 2008). In addition, the Brγ line from hot hydrogen gas is clearly seen, and is shown to arise in a region that is smaller than (interior to) the hot inner dust that produces most of the continuum emission. Brγ lines originate in very hot gas (~ 10000 K) and similar results have been found using molecular lines (CO, water vapor) that trace more relevant disk regions (Tatulli et al. 2008, Eisner et al. 2009).

In the next decade, instruments that already exist will be able to expand beyond these pioneering efforts by providing detailed spectro-interferometric measurements for a large sample of disk objects across the stellar mass range. These new measurements will allow us to elucidate important inner disk phenomena that directly confront the how and when of planet formation:

***Detailed Structure of the Dust Evaporation Front:*** Spatially resolved measurements constrain the nature (geometry and composition) of the dust evaporation front. This knowledge is crucial because its shadow extends to the inner few AU (e.g., Dullemond 2002), influencing the properties of the terrestrial planet formation zone (e.g. reducing the amount of starlight that reaches it, including disk ionization by stellar magnetosphere X-rays). Mid-infrared observations can provide critical input to models of dust-phase mineralogy. In combination with laboratory experiments (e.g., Jaeger et al. 1998) we can address the evolution of dust from which Earth-like planets are built (for example, the poorly understood origin of the crystalline nature of the inner dust, van Boekel et al. 2005). Also, current models ignore the effect on the inner dust rim from heating and opacity by gas interior to the dust sublimation radius, a complete physical description demands that both are treated self-consistently.

***Density Profiles and Planet Formation:*** Gas giant planet formation depends on the gas and dust columns during the optically-thick disk stage. Measurements of the surface brightness constrain the disk density and temperature profiles. By performing these measurements for young stars with a range of masses (from T Tauri to Herbig Ae/Be), it becomes possible to distinguish between the gravitational collapse and core accretions modes of planet formation. Indeed, over 300 exoplanets have been discovered so far: will we be able to explain the observed differences in planetary architectures as differences in disk conditions? Another important application is the study of disk "dead zones": disk regions predicted to be poorly ionized and therefore where the magneto-rotational instability, source of disk viscosity and accretion, is inactive. Dead zones, if they exist, can have important implications for planet formation: are they the regions in which planets form? (e.g. Glassgold et al. 1997); do they save planetary systems by halting migration? (e.g., Matsumura et al. 2007). Observationally, because they can accumulate large local densities without accreting onto the central star, dead zones clearly can radically change the disk properties at specific radii. In fact, mass accumulation in dead zones followed by gravitational instability has been proposed as an explanation for the outbursts observed in FU Orionis objects (e.g., Zhu et al. 2008). These systems have long been recognized as powerful laboratories for studying the accretion process itself (Malbet et al. 1998, 2005) and understanding FU Ori outbursts as they proceed may provide the best evidence for the existence of dead zones and for studying how they affect planet formation.

***The Star – Disk Connection:*** Accretion disk models have tended to focus on two separate scales: **(i)** the process of mass and angular momentum in the outer disk (10s-100s AU), and **(ii)** gas accretion from a few $R_*$ on to the stellar photosphere. Processes involving interactions of the disk with the star are poorly understood (e.g., disk locking and winds, Matt & Pudritz 2005, 2008), yet are crucial to the formation of the young star and its circumstellar disk. Also, despite their ubiquity and importance, jets in young stellar objects are poorly understood, as are the timescales for their bursty nature. By measuring the inner gas disk temperature and density profiles, for regions from 1/10s AU to the stellar surface, we can help clarify the "star – disk connection" and study the mechanisms responsible for angular momentum transport. The removal of angular momentum is key to understanding the origins of the planets, since it determines the rate at which material spirals in to accrete on the young star. Modelling Balmer line profiles using hybrid disk – wind models can provide the disk accretion rates. Combined with new measurements of inner gas surface density, we can then directly constrain disk viscosity parameter ($\alpha$) and probe models of transport. Disk viscosity also controls the migration speed, a crucial process not only in shaping planetary architectures, but also possibly in determining the water content of terrestrial planets, delivered by inward migration of icy planetesimals (Cyr et al. 1998, Drake 2005).

## Direct Imaging of the Inner Disk and Proto-Planets

Without imaging, we rely on educated guesses to model and interpret the observations, photometric, spectroscopic as well as interferometric. In this case our understanding is only as secure as the completeness of our models and, as we described above, there are clear signs that we are still missing key physical ingredients. Milliarcsecond imaging can be expected to become commonplace during the next decade, providing a powerful new tool with which to confront our basic paradigms, and the likely complexity of disk morphologies.

Just as direct imaging at infrared to mm wavelengths has led to fundamentally-new insight into the structure of debris disks (e.g. Weinberger et al. 1999, Koerner et al. 1998, Jayawardhana et al. 1998, Wilner et al. 2002) and their link to dynamical interactions with forming planets (e.g. Takeuchi & Artymowicz 2001, Kenyon & Bromley 2002, Mayer et al. 2002), interferometric imaging of the more massive, less evolved disks promises to reveal the earliest stages of planet formation. As a prelude, images of the hot dust around two nearby massive pre-main sequence stars, LkHα 101 and MWC 349, have already been made using aperture masking interferometry (Tuthill et al. 2001, Danchi et al. 2001). In fact, the central "hole" of flux seen in the LkHα 101 image (Figure 3) led to the original speculation that young stellar disks are harboring much larger dust-free inner cavities than expected. Moreover, monitoring of the disk emission has detected unexpected changes (Tuthill et al. 2002) over a few-year period, holding the tantalizing possibility that we are seeing large-scale dust structures in orbital motion in the disk.

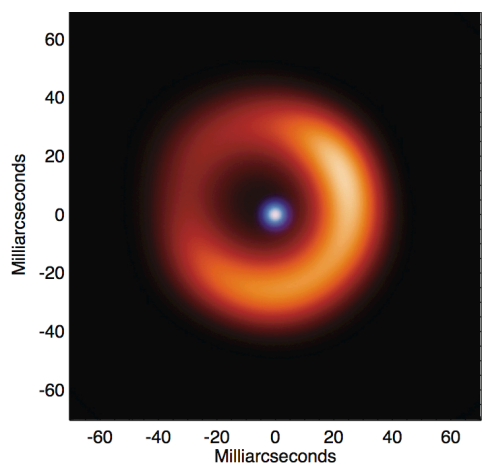

**Figure 3:** Image of the massive young stellar object LkHα 101 formed by aperture masking of a single Keck telescope (Tuthill et al. 2001). The asymmetric brightness seen in the circumstellar ring of emission (hot dust) is likely due to flaring of the disk seen at a slight inclination. This image illustrates what is achievable by long baseline interferometric imaging, but for more typical objects, and with ×30 better angular resolution.

More hints at what is to come are provided by the prototypical Herbig Ae star AB Aur. It has a larger disk and is easier to observe today than the typical T Tauri star, and shows "spiral" or arc structures in the outer CO disk (Corder et al. 2005) and scattered light (Grady et al. 1999); a possible "gap" at 100 AU, possibly due to a forming planet, based on coronagraphy (Oppenheimer et al 2008); a sudden jump in the abundance of small grains within 7 AU deduced from its temperature profile, possibly due to planetesimal collisions (Tannirkulam et al 2008);

and even evidence from infrared interferometry for a disk hotspot within a few AU, perhaps from a circumplanetary accretion disk (Millan-Gabet et al. 2006).

Other young stars have shown signs of inner disk structures: the distinct signature of a disk hotspot has been detected in the FU Orionis system (Malbet et al. 2005); MWC 275 has displayed strong photometric brightening of the inner disk (Sitko et al. 2007), perhaps related to its episodic microjets; "pre-transitional" disks have been found showing both hot dust emission and outer disk emission, suggesting a gap clearing (Espaillat et al. 2008); and recent spectroastrometric results (Pontoppidan et al. 2008) indicate that there are significant deviations from Keplerian rotation in some systems and significant azimuthal asymmetries are sometimes present, perhaps as a result of planet formation processes. While each of these measurements has some uncertainties, they force fascinating questions: Are we seeing a new kind of disk instability, or evidence for accreting proto-planets embedded in young disks? With model-independent imaging we can approach these questions with a powerful new tool, by searching for large scale inhomogeneities in the disk emission. The range of phenomena already observed promises transformative rewards from these investigations for years to come.

Through infrared surveys of multiple star forming regions, we have learned that the probability of a young star having an optically-thick disk drastically falls as a function of age, with a time scale of about three million years (Haisch et al. 2001). The process that destroys the disk is also very fast (100,000 years), given the observed rarity of "transitional" disk systems that still show signs of on-going accretion and large few AU dust gaps. Giant planet migration could also be very efficient compared to disk lifetimes (Matsumura et al. 2007).

While planet formation might explain these timescales, this discussion exposes much about what we don't know. When do planets really form in these systems? Are there multiple generations of planets that interact with the gaseous disk and migrate inward only to fall onto the star? Or is there one planet formation episode that uses most of the inner disk material and leaves the remnant gas to be rapidly photo-evaporated away? Especially in the former case, we can expect lots of structure to be observed in the inner 10 AU. Mm-wave interferometers like ALMA will be able to probe these signatures in the outer disk beyond 10 AU, while infrared interferometers are needed to image the inner disks.

For some transition disks, already one can persuasively argue that the simultaneous presence of on-going accretion and the large dust gaps deduced from the spectral energy distribution argue against "caught-in-the-act" photo-evaporation and strongly suggests gap formation by a Jovian mass planet (Najita et al. 2007b). Depending on the system, these gaps and the Jovian planets themselves could be observed directly with capabilities that will become available in the next decade, including nulling interferometry and coronography on a 20 - 30 m class telescope.

Continuing beyond the T Tauri phase and toward later evolutionary stages, we note important complementary approaches towards providing strong constraints on planet formation mechanisms, described in detail elsewhere: structures of debris disks (e.g., Kuchner & Holman 2003, Chiang et al. 2009); the measurement of binary orbits and precise masses for young stars (e.g., Boden et al. 2007), with which to calibrate stellar evolution models; and the astrometric or direct detection of fully formed planets around young stars (1 − 100 Myr, Beichman 2001, Marois et al. 2008).

# Conclusions

The tremendous diversity in the properties of the > 300 exoplanets discovered to date (from hot-Jupiters to hot-Earths) highlights the need for observational probes of inner disks that may guide the development of a predictive theory of the formation and evolution of planetary systems.

Current infrared facilities in the US collectively offer angular resolution as fine as < 1 mas, simultaneously with spectral resolution (R ~ few 1000), and have demonstrated the ability to image complex morphologies (Monnier et al. 2007, Zhao et al. 2008). In the context of disk studies, developments under way are aimed primarily at increasing the sensitivity, and realistic projections imply sample sizes of hundreds of young stellar objects across the mass range, from which to derive meaningful statistics and trends.

Observations that can spatially resolve the inner disk will offer unique opportunities to directly probe the physical conditions in young disks. In combination with complementary techniques such as high-resolution spectroscopy, mm-wave imaging of the outer disk, extreme adaptive optics with coronagraphy, and advances in modeling the relevant phenomena, this coming decade promises to transform our understanding of how, when and where planets form around other stars.

# References

Beichman C. A. 2001, in ASP Conf. Ser., 244, 376
Boden A. et al. 2007, ApJ, 670, 1214
Chiang E. et al. 2009, ApJ, in press
Corder S. et al. 2005, ApJ, 622L, 133
Cyr K. et al. 1998, Icarus, 135, 537
Danchi W. et al. 2001, ApJ, 562, 440
Drake M. J. 2005, M&PS, 40, 519
Eisner J. 2007, Nature, 447, 562
Eisner J. et al. 2009, ApJ, in press
Espaillat C. et al. 2008, ApJL, 682, L125
Glassgold A. et al. 1997, ApJ, 480, 344
Grady C. et al. 1999, ApJL, 523, L151
Haisch K. E. et al. 2001, ApJ, 553L, 153
Jaeger C. et al. 1998, A&A, 339, 904
Jayawardhana R. et al. 1998, ApJL, 503, L79
Kenyon, S. J. & Bromley B. C. 2002, ApJ, 577L, 35
Koerner D. et al. 1998, ApJL, 503, L83
Kuchner M. & Holman M. 2003, ApJ, 588, 1110
Kraus S. et al. 2008, ApJ, 676, 490
Malbet F. et al. 1998, ApJ, 507, 149
Malbet F. et al. 2005, A&A, 437, 672
Marois C. et al. 2008, Science, 322, 1348
Matt S. & Pudritz R. 2005, ApJ, 632, 135
Matt S. & Pudritz R. 2008, ApJ, 678, 1109
Matsumura S. et al. 2007, ApJ, 660, 1609
Mayer L. et al. 2002, Science, 298, 1756
Millan-Gabet R. et al. 2006, ApJ, 645L, 77
Millan-Gabet R. et al. 2007, in Protostars & Planets V, University of Arizona Press, Tucson
Monnier J. D. et al. 2007, Science, 317, 342
Najita J. et al. 2007a, in Protostars & Planets V, University of Arizona Press, Tucson
Najita J. et al. 2007b, MNRAS, 378, 369
Oppenhemier B. R. et al. 2008, ApJ, 679, 1574
Pontoppidan K. M. et al. 2008, ApJ, 684, 1323
Shakura N. I. & Syunyaev R. A 1973, A&A, 24, 337
Sitko M. et al. 2008, ApJ, 678, 1070
Takeuchi, Taku; Artymowicz, P. 2001, ApJ, 557, 990
Tannirkulam A. et al. 2008, ApJ, 689, 513
Tatulli E. et al. 2008, A&A, 489, 1151
Tuthill P. G. et al. 2001, Nature, 409, 1012
Tuthill P. G. et al. 2002, ApJ, 577, 826
van Boekel R. et al. 2005, A&A, 437, 189
Weinberger A. et al. 1999, ApJL, 525, L53
Wilner D. et al. 2002, ApJL, 569, L115
Zhao M. et al. 2008, ApJL, 684, L95
Zhu Z. et al. 2008, ApJ, in press